\documentclass[fleqn,usenatbib]{mnras}

\usepackage{newtxtext,newtxmath}

\usepackage[T1]{fontenc}
\usepackage{ae,aecompl}

\usepackage{graphicx}	
\usepackage{amsmath}	
\usepackage{amssymb}	
\usepackage{nicefrac}
\usepackage{ulem}
\usepackage{color}

\usepackage{tikz, graphics} 
\usetikzlibrary{arrows.meta, 
                bending,     
                patterns     
               }

\newcommand{\cutt}[1]{\textcolor{blue}{}}

\newcommand{\msun}{\ensuremath{\mathrm{M}_{\odot}}}
\newcommand{\kmsun}{\ensuremath{\mathrm{kM}_{\odot}}}
\newcommand{\Myr}{\ensuremath{\mathrm{Myr}}}
\newcommand{\kyr}{\ensuremath{\mathrm{kyr}}}
\newcommand{\K}{\ensuremath{\mathrm{K}}}
\newcommand{\kK}{\ensuremath{\mathrm{kK}}}
\newcommand{\yr}{\ensuremath{\mathrm{yr}}}

\title[The lives and deaths of supermassive stars]{On Monolithic Supermassive Stars}
\author[T. E. Woods et al.]{
Tyrone E. Woods,$^{1,2}$\thanks{Plaskett Fellow}\thanks{E-mail: tyrone.woods@nrc-cnrc.gc.ca}
Alexander Heger,$^{2-6}$
Lionel Haemmerl\'e$^{7}$
\\
$^{1}$National Research Council of Canada, Herzberg Astronomy \& Astrophysics Research Centre,\\ 5071 West Saanich Road, Victoria, BC V9E 2E7, Canada\\
$^{2}$School of Physics \& Astronomy, Monash University, Clayton 3800, Victoria, Australia\\
$^{3}$Joint Institute for Nuclear Astrophysics, 1 Cyclotron Laboratory, National Superconducting Cyclotron Laboratory,\\ Michigan State University, East Lansing, MI 48824-1321, USA\\
$^{4}$Tsung-Dao Lee Institute, Shanghai 200240, China\\
$^{5}$Center of Excellence for Astrophysics in Three Dimensions (ASTRO-3D), Australia\\
$^{6}$Australian Research Council Centre of Excellence for Gravitational Wave Discovery, Clayton, VIC 3800, Australia\\
$^{7}$D{\' e}partement d'Astronomie, Universit{\' e} de Gen{\` e}ve, Chemin des Maillettes 51, 1290 Versoix, Switzerland\\
}

\date{Accepted XXX. Received YYY; in original form ZZZ}

\pubyear{2020}

\begin{document}
\label{firstpage}
\pagerange{\pageref{firstpage}--\pageref{lastpage}}
\maketitle

\begin{abstract}
Supermassive stars have been proposed as the progenitors of the massive ($\sim 10^{9}\,\msun$) quasars observed at $z\sim7$.  Prospects for directly detecting supermassive stars with next-generation facilities depend critically on their intrinsic lifetimes, as well as their formation rates.  We use the 1D stellar evolution code \textsc{Kepler} to explore the theoretical limiting case of 
zero-metallicity, non-rotating stars, formed monolithically with initial masses between $10\,\kmsun$ and $190\,\kmsun$. 
We find that stars born with masses between $\sim60\,\kmsun$ and $\sim150\,\kmsun$ collapse at the end of the main sequence, burning stably for $\sim1.5\,\Myr$.  More massive stars collapse directly through the general relativistic instability after only a thermal timescale of $\sim3\,\kyr$--$4\,\kyr$. 
The expected difficulty in producing such massive, thermally-relaxed objects,
together with recent results for currently preferred rapidly-accreting formation models, 
suggests that such ``truly direct'' or ``dark'' collapses may not be typical for supermassive objects in the early Universe. 
We close by discussing the evolution of supermassive stars in the broader context of massive primordial stellar evolution and the possibility of supermassive stellar explosions.

\end{abstract}




\section{Introduction}

The detection of high redshift quasars which have reached $\sim 10^{9}\,\msun$ within the first billion years presents a stark challenge to our understanding of the evolution of structure in the Universe \citep[e.g.,][]{HL01,Mortlock11}.  A Population III stellar-mass black hole ($\sim100\,\msun$) forming at $z\sim20$--$30$ would need to continuously accrete at the Eddington rate in order to reach this mass by $z\sim7$, an unlikely event given the strong ionizing feedback from such an object and the gas reservoir typically available in their natal environments \citep{WF12}.  In light of these difficulties, alternative mechanisms for producing more massive initial black hole masses (``seeds'') have been increasingly favoured, with the collapse of supermassive ($\sim 10^{5}\,\msun$) stars emerging as a promising alternative scenario for the progenitors of massive black holes in the early Universe \citep[e.g.,][]{Rees84,Volonteri10,Woods19}. 

The existence of supermassive stars, with masses $>10^{5}\,\msun$, was first suggested as a possible explanation for the nature of luminous, extragalactic radio sources \citep{HF63,Iben63}.  Although these are now understood to be accreting supermassive black holes, a number of pathways leading to the formation of such objects via an initially supermassive star remain viable \citep{Rees84}. Classically, perhaps the most commonly-invoked route to producing such an object has been via the rapid {\it collapse of a dense stellar cluster} \citep[e.g.,][]{BR78}, although theoretical models have thus far found it difficult to produce seeds more massive than $\sim 3$--$30\times 10^{3}M_{\odot}$ in this way. More recently, a number of channels have also been proposed for massive seed formation via extremely rapid accretion onto a single (or small number of) seeds.  In the atomic-cooling halo scenario \citep[e.g.,][]{BL03,Dijkstra08,Regan17}, photo-destruction of the molecular hydrogen in a primordial halo at $z\sim10$--$20$ by an intense Lyman-Werner flux permits the halo gas to reach $\sim8\,\kK$ before radiative losses from atomic hydrogen line transitions can lock in the temperature, bringing the Jeans mass and typical infall rates up to $\sim100\,\kmsun$ and $0.1$--$10\,\msun\,\yr^{-1}$.  The fates of the resulting objects have been examined in increasing detail in the last few years \citep[e.g.,][]{Hosokawa13,Woods17,Haemmerle18b}.  Large baryonic streaming velocities may be able to assist in suppressing star formation in the process, and may even be sufficient in the absence of a substantial Lyman-Werner flux \citep[e.g.,][]{TLH13}.  Alternatively, the merger of massive gas-rich protogalaxies may be able to create the necessary conditions for an unstable, massive nuclear disk to drive extremely rapid inflows ($\sim100$--$1000\,\msun\,\yr^{-1}$), providing another channel for the formation of even more massive seeds \citep{Mayer10}.

There is growing indirect evidence for the existence of massive seeds, and perhaps supermassive stars, e.g., the apparently low critical Lyman-Werner flux needed for atomic-cooling halos to form \citep{Wise2019}, and the low occupation fraction of IMBHs in dwarf galaxies \citep{Mezcua2019}.  None of the scenarios listed above,  however, are without difficulties \citep[ e.g.,][]{Volonteri10,Latif16,Woods19}.  Recent efforts have examined the evolution and fates of the growing central objects in the rapidly-accreting (e.g., atomic-cooling halo) as well as protogalactic merger scenarios 
\citep{Woods17,Woods19,Haemmerle18a,Haemmerle2019}.  Here, we investigate the fates of ``monolithically-formed'' supermassive stars as an idealized case in the limit of wholly thermally-relaxed objects.
Previously, \cite{Fuller86} carried out the first early investigation into the evolution and fate of supermassive stars to use the \textsc{Kepler} stellar evolution code \citep[though see also][]{AF72a,AF72b}.  Here, we provide a modern update incorporating an adaptive nuclear burning network, as well as significant refinement of their results for zero-metallicity supermassive stars, in order to determine their ultimate fates.

In Section \ref{SMS}, we review the properties of supermassive stars before outlining the numerical method used in our simulations.  In particular, we note the key upgrades which have been made to the stellar evolution and hydrodynamics code {\sc Kepler} since the seminal work of \cite{Fuller86}, notably the inclusion of an adaptive network allowing for a complete treatment of all relevant nuclear reactions \citep{Woosley04}.  In Section \ref{results}, we delineate for the first time the boundaries as a function of stellar mass for stable helium-burning, stable hydrogen burning, and ``truly direct'' collapse to a black hole (i.e., without a prior phase of hydrostatic nuclear-burning). 
We provide a summary of the evolutionary fates of the most massive stars in Section~\ref{results}.  We then close in section \ref{discussion} with a brief discussion of these results in a broader context. 

\section{Modelling Supermassive stars}\label{SMS}

\subsection{Analytical Estimates}

The first investigations of the plausible physical characteristics of supermassive stars were carried out by \cite{HF63} and \cite{Iben63}. These early studies ignored any details of the formation of the $10^4$--$10^8\,\msun$ objects which they considered, implicitly assuming them to have formed instantaneously, though not in a thermally-relaxed state.  Throughout this work, we shall refer to such models as ``monolithic'' supermassive stars, in contrast with supermassive stars formed through extremely rapid accretion in massive, atomically-cooled halos \citep[e.g.,][]{Regan17}, for which there are striking differences in the internal structure \citep[e.g.,][]{Hosokawa13}.  A high-entropy, monolithic model is commonly invoked as a reasonable approximation for a supermassive star formed in the collapse and coalescence of a dense stellar cluster \citep[e.g.,][]{BR78,Fuller86,Denissenkov14,Gieles18}. 
More generally, however, such models are instructive as a limiting case, being the most compact, thermally-relaxed means of constructing a model supermassive star.
In this section, we focus only on what is essential to grasp the physics determining the fate of primordial supermassive stars as a function of their initial mass.  For a more thorough review, see \cite{Fuller86} and \cite{Woods19}.

Radiation pressure easily provides the dominant contribution to maintaining hydrostatic equilibrium in supermassive stars.  Therefore, in the case of monolithically-formed supermassive stars their structure is well-approximated as a polytrope of index $n=3$.  The ratio of gas-to-total pressure is thus given by

\begin{equation}
\beta = \frac{P_{\rm{gas}}}{P_{\rm{tot}}} \approx \frac{4.3}{\mu}\left(\frac{\rm{M}_{\odot}}{\rm{M}}\right)^{\nicefrac12}
\end{equation}

\noindent \citep{Chandrasekhar39}, where $\mu$ is the mean molecular weight and M is the stellar mass. The local adiabatic index within such a star is everywhere very nearly $\nicefrac43$

\begin{equation}\label{loc_ad}
\Gamma _{1} \approx \frac{4}{3} + \frac{\beta}{6}
\end{equation}

\noindent with gas pressure providing only a small additional perturbation.  A classical polytropic star with $\Gamma _1 = \nicefrac43$ has no natural length scale, and zero total energy; such objects are unstable and may be expected to explode or collapse on their dynamical timescale. \cite{Chandrasekhar64} showed that a fully general relativistic treatment yields an instability at a slightly greater adiabatic index of

\begin{equation}
\Gamma _{\rm{crit}} \approx \frac{4}{3} + K\left(\frac{2GM}{Rc^2}\right)
\end{equation}

\noindent where for an n = 3 polytrope, $K \approx 1.12$.  Therefore, recalling Eq.~(\ref{loc_ad}), we may expect the general relativistic instability to induce the dynamical collapse of any supermassive star above some critical mass, depending also on its radius and mean molecular weight.  In practise, this limit has more commonly been recast using more convenient variables, such as a limiting critical central density:

\begin{equation}
\rho _{\rm{crit}} \approx 2\times 10^{18}\left(\frac{0.5}{\mu}\right)^3\left(\frac{M_{\odot}}{M}\right)^{\nicefrac72} \rm{g}\,\rm{cm}^{-3}
\end{equation}

\noindent Whether and when a supermassive star undergoes a collapse induced by the general-relativistic instability, however, also depends on its thermal and nuclear evolution.

The luminosity of any very massive radiation-dominated star will be very nearly at the Eddington limit:

\begin{equation}
L_{\rm{Edd}} = 4 \pi G M c\,\kappa^{-1}    
\end{equation}

\noindent with electron scattering the dominant source of opacity:

\begin{equation}
\kappa = \frac{\sigma_{\rm{T}}}{\rm{m}_{\rm{p}}}\left(X + \frac{1}{2}Y\right)
\end{equation}

\noindent where X and Y are the hydrogen and helium abundances, $\sigma_{\rm{T}}$ is the Thomson cross section, and $\rm{m}_{\rm{p}}$ is the proton mass. Here $\kappa$ varies from $0.825\,\sigma_{\rm{T}}/\rm{m}_{\rm{p}}$ to $0.5\,\sigma_{\rm{T}}/\rm{m}_{\rm{p}}$ for primordial composition and pure helium, respectively. The total energy available from nuclear-burning of hydrogen is $Q\approx\epsilon\,M$, where $\epsilon \approx 6.4\times 10^{18}\,\mathrm{erg}\,\mathrm{g}^{-1}$.  This gives us an estimate of the hydrogen-burning main sequence lifetime:

\begin{equation}
\tau _{\rm{nuc}} = \frac{\kappa \epsilon}{4\pi G c} \approx 1.6\,\Myr\label{taunuc}
\end{equation}

\noindent independent of the mass. Should the star undergo stable helium-burning prior to collapse, its lifetime can exceed this by $\sim10\,\%$. 

In the absence of nuclear burning, or before its onset in the event that the supermassive star is initially formed in some ``puffed-up'' high-entropy state, the star will contract on its thermal timescale until reaching a critical energy/central density \citep{ShapiroTeukolsky1983}:

\begin{equation}
\tau _{\rm{th}} = \frac{|\rm{E}_{\rm{crit}}|}{\rm{L}_{\rm{Edd}}} \sim 6,\!300\,\yr\,\left(\frac{M}{10^5M_{\odot}}\right)^{-1}\label{tauth}
\end{equation}

For a ``pre-main sequence,'' primordial-composition supermassive star undergoing thermal relaxation, nuclear-burning of hydrogen can initially proceed only through the PP III chain once the core of the contracting star reaches temperatures of $\gtrsim 2.3\times 10^{7}\,\K$.  The rate of energy release through this reaction alone, however, is insufficient to halt the collapse of the massive, primordial-composition star \citep[e.g.,][]{Fuller86,Begelman10}.  Instead, the contraction continues until triple-$\alpha$ burning of helium sets in at $\rm{T}_{\rm{c}}\gtrsim 10^{8}\,\K$.  The fate of monolithic supermassive stars with masses $\gtrsim 10^{5}\,\msun$, then, becomes a question of whether triple-$\alpha$ burning can produce a sufficient CNO abundance to catalyze hydrogen burning before the star has contracted to the point of satisfying the general relativistic instability.

\subsection{Simulating supermassive stars with Kepler}

In order to quantitatively delineate the boundary between supermassive stars which survive to nuclear-burning, and those which undergo a truly direct collapse, we must carry out detailed stellar evolutionary calculations.
In this work, we carry out our analysis using the 1-dimensional lagrangian hydrodynamics and stellar evolution code {\sc kepler} \citep{WZW78,WHW02}. {\sc kepler} solves for the conservation of angular momentum and energy, including acceleration and viscosity terms, as well as first-order (post-Newtonian) corrections to gravity:

\begin{equation}\label{cons_mom}
\frac{dv}{dt} = 4\pi r^{2} \frac{\partial P}{\partial m_{r}} - \frac{G_{\rm{rel}}m_{r}}{r^{2}} + \frac{4\pi}{r}\frac{\partial Q}{\partial m_{r}}
\end{equation}

\begin{equation}\label{cons_en}
\frac{du}{dt} = -4\pi P \frac{\partial}{\partial m_{r}} (vr^{2}) + 4\pi Q \frac{\partial}{\partial m_{r}}\left(\frac{v}{r}\right) - \frac{\partial L}{\partial m_{r}} + \epsilon
\end{equation}

\noindent where in eq. \ref{cons_mom}, the acceleration in any Lagrangian zone is found from the sum of the pressure gradient, gravitational acceleration, and viscous drag, while the net energy flux for the same is found from the sum of the work, viscous dissipation, radiative flux, and energy generation. The post-Newtonian correction is included through the modified gravitational constant:

\begin{equation}
G_{\rm{rel}} = G \left(1 + \frac{P}{\rho c^{2}} + \frac{4\pi Pr^{3}}{m_{r}c^{2}}\right)\left(1 - \frac{2Gm_{r}}{rc^{2}}\right)^{-1}
\end{equation}

\noindent while the viscous terms include the factor:

\begin{equation}
Q = \frac{4}{3}\,\eta_{\nu}r^4\frac{\partial\left(\frac{v}{r}\right)}{\partial r}
\end{equation}

\noindent with $\eta _{\nu}$ the dynamic viscosity as given in \cite{WZW78}, including both the real and artificial viscosity. The latter allows us to damp out acoustic oscillations during quiescent periods of the evolution of a model.  Nuclear-burning is included using a full adaptive network implicitly coupled to the hydrodynamics \citep{Woosley04}.  Note that this is a dramatic improvement over the 10-isotope APPROX network \citep{WW81} used in the original study of \cite{Fuller86}. {\sc kepler} uses a Helmholtz-like equation of state \citep{TS00} including electron-positron pair production; relativistic and non-relativistic, degenerate and non-degenerate electrons; and radiation. Convection is carried out in a time-dependent manner as described in \cite{WZW78}.  Convective heat transport is included only when a zone is convective by the Ledoux criterion. For accreting models, accretion is carried out as described in \cite{Woosley04} and \cite{Woods17}.

{\sc Kepler} partitions each supermassive star model into an arbitrarily large number of zones; here we typically resolve $\sim1$--$2,\!000$ zones, with a particularly fine mesh near both the centre ($\sim 10^{30}\,\mathrm{g}$ per zone) and the surface (down to $\sim 10^{26}\,\mathrm{g}$ per zone). For each model step, {\sc Kepler} makes an arbitrarily large initial guess for the duration of the next time step, and then scans through all zones in order to find the shortest time step needed in order to avoid too great a change in the fractional radius, temperature, luminosity, or density of any zone per time step, and to follow the emergence of any shocks.  In this way, we can follow the long-term evolution of the star over thermal and nuclear timescales, dropping to sufficiently short steps so as to resolve hydrodynamic timescales only at the emergence of an instability.

\begin{figure*}
\begin{tabular}{cc}
\includegraphics[width=0.5\textwidth]{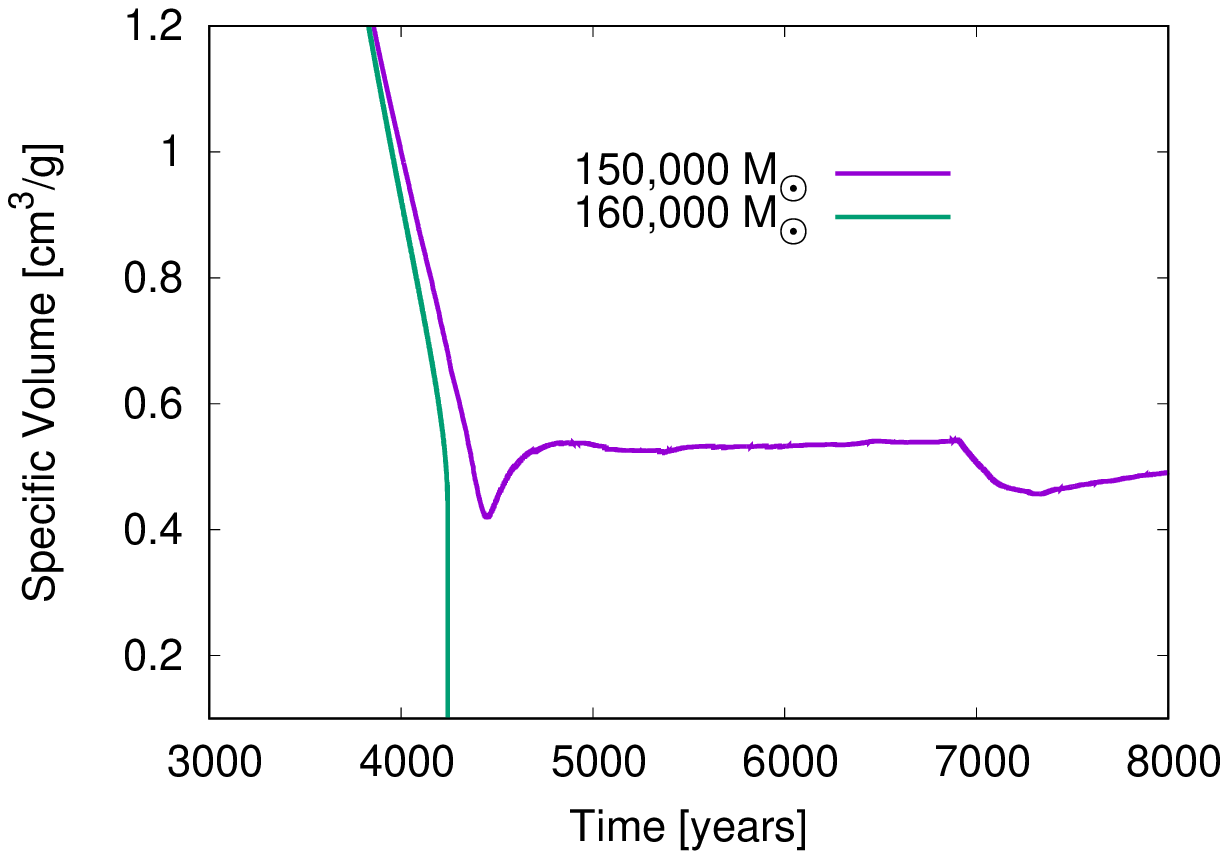} &
\includegraphics[width=0.5\textwidth]{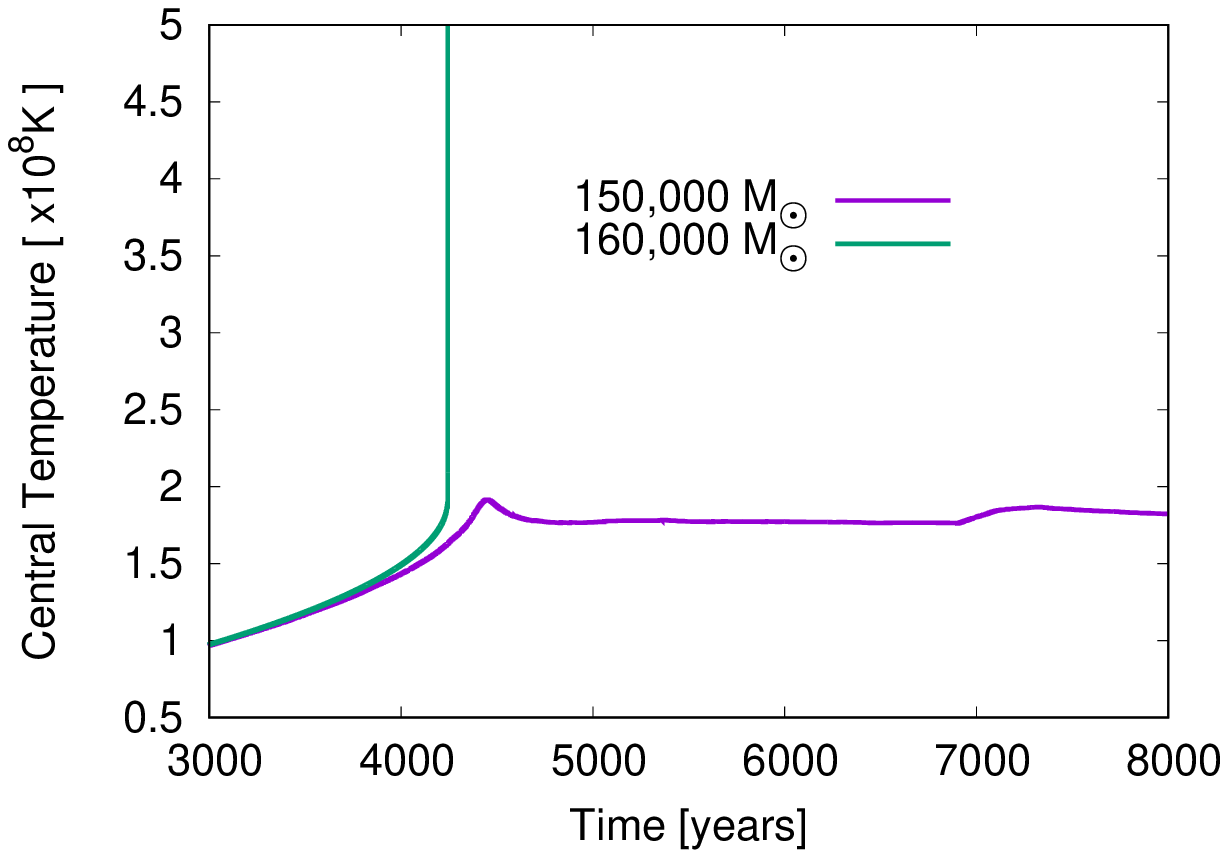} \\
\end{tabular}
\begin{center}
\caption{Time evolution of the central specific volume (left panel) and central temperature (right panel) after approximately $\tau_{\rm{KH}}$ for monolithic SMSs of $150\,\kmsun$ (purple) and $160\,\kmsun$ (green). }\label{central}
\end{center}
\end{figure*}

Each model is initialized as a $n=3$ polytrope of a given mass, with the initial entropy then set by the choice of central density. Here we choose $\rho_{\rm{c,i}} = 0.001\,\rm{g}\,\rm{cm}^{-3}$ in order to simulate relatively high entropy protostars formed from the rapid collapse of a dense stellar cluster, as in \cite{Fuller86}.  We assume primordial composition consistent with Big Bang Nucleosynthesis with a baryon-to-photon ratio of $\sim 6\times 10^{-10}$, with initial mass fractions of H, $^{4}$He, $^{3}$He, $^{2}$H, and Li of $\sim0.75$, $0.25$, $2.1\times10^{-5}$, $4.3\times 10^{-5}$, and $1.9\times 10^{-9}$, respectively \citep{cyb01,cyb02}.  We ignore mass losses both due to winds \citep[expected to be negligible, see e.g.,][]{Vink01} and pulsations \citep{baraffe01}.  We model protostars with initial masses between $30\,\msun$ and $190\,\kmsun$ with a resolution of $10\,\kmsun$.  In order to carefully resolve the onset of nuclear-burning, we limit the time step to $10^{6}$--$10^{7}\,\mathrm{s}$ during the initial hydrostatic contraction/thermal relaxation of our models (i.e., the first $\sim 2\times 10^{11}\,\mathrm{s}$).

\section{Results}\label{results}

The fate of each model after approximately a thermal timescale depends on whether the energy released from nuclear burning can halt contraction before the onset of a runaway dynamical collapse, arising due to the post-Newtonian Chandrasekhar instability. We find a transition between hydrostatic hydrogen-burning and dynamical collapse occurs above  $\gtrsim 150\,\kmsun$. This is illustrated in Fig.~\ref{central}.  The central specific volume ($1/\rho_{\rm{c}}$) and central temperature ($\rm{T}_{\rm{c}}$) are plotted as a function of time, for either end of this mass range, after approximately one thermal relaxation time has elapsed.  For objects of $160\,\kmsun$ (and greater), we find that nuclear-burning is unable to halt contraction and ultimately collapse.  We terminate our calculations once the central temperature has exceeded $1.5\times 10^{10}\,\K$, at which point we find infall velocities have reached a few~\% of the speed of light.  Although our post-Newtonian approximation is no longer valid beyond this point, such objects are well above the pair instability limit for direct black hole formation, and therefore may be expected to continue to collapse to form black holes \citep{Heger02}. 

Objects formed with initial masses of $150\,\kmsun$ and below are found to survive to hydrostatic nuclear-burning; these are the most ``supermassive'' objects which may be called ``stars.'' 
In this mass range, the onset of triple-$\alpha$ helium-burning once central temperatures have reached $\gtrsim 10^{8}$K is able to produce sufficient CNO to catalyze hydrogen-burning before the contraction becomes unstoppable. Interestingly, at the upper bound of this regime, the contraction is reversed at somewhat greater densities than predicted to be stable against the post-Newtonian instability in the polytropic approximation. This is reasonable; polytropic models are unable to account for the energy release from nuclear-burning, which reaches a peak at a central density of $\sim 2\times \rho _{\rm{crit}}$ for our $150\,\kmsun$ model before reversing the collapse. This is seen in Fig.~\ref{central} as a turn-around in the central density and temperature. It leaves the star in a slightly inflated state, which persists for $\sim2,\!500\,\yr$, comparable to the thermal timescale, before the star settles onto the hydrogen-burning main sequence. Note that during this time we find significant noise in the central CNO abundances, resolved only once central convection is firmly established after $\sim$ 20,000 years. 

\begin{figure}
\begin{center}
\includegraphics[width=0.5\textwidth]{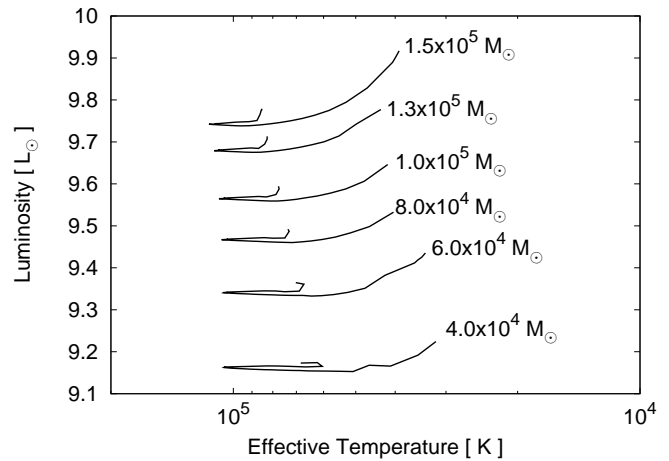}
\caption{HR diagram showing the evolution of a representative subset of our simulations which reach the main sequence, from shortly after their onset up until the end of the hydrogen-burning main sequence ($\sim1.5\,\Myr$).}\label{HRdiagram}
\end{center}
\end{figure}

After this reversal, our models with $M\leq150\,\kmsun$ enter the main sequence with central temperatures of $\sim1.8\times10^{8}\,\K$, effective temperatures of $\sim$1--1.1$\times10^{5}\,\K$, and with luminosities which scale linearly with the total mass, from $\sim(1$--$6)\times 10^{9}\,\mathrm{L}_\odot$ for $30\,\kmsun$--$150\,\kmsun$ (see also Table~\ref{tab:MS}).  These ZAMS effective temperatures and luminosities are in very close agreement with that for a nearly Eddington-luminosity, $n=3$ polytrope assuming radiation pressure support only \citep[e.g.,][]{HF63}, confirming the consistency of our simulations.  The subsequent evolution of a representative selection of our models up until the end of the main sequence is shown in Fig.~\ref{HRdiagram}.

\begin{table}
    \centering
    \begin{tabular}{c|c|c|c}
        M &  $\rm{L}_{\rm{ZAMS}}$ & $\rm{T}_{\rm{eff,ZAMS}}$ & $\rm{T}_{\rm{c,ZAMS}}$\\ $\rho_{\rm{c,ZAMS}}$\\
        ($\kmsun$) &  (GL$_\odot$) & (MK) & (GK) \\
        \hline
         30 & 1.09 & 0.104 & 0.177\\
         40 & 1.48 & 0.105 & 0.179\\
         50 & 1.85 & 0.104 & 0.180\\
         60 & 2.24 & 0.106 & 0.176\\
         70 & 2.60 & 0.106 & 0.177\\
         80 & 2.86 & 0.107 & 0.178\\
         90 & 3.39 & 0.108 & 0.179\\
        100 & 3.65 & 0.108 & 0.180\\
        110 & 4.17 & 0.109 & 0.180\\
        120 & 4.43 & 0.111 & 0.181\\
        130 & 4.95 & 0.111 & 0.180\\
        140 & 5.21 & 0.114 & 0.187\\
        150 & 5.73 & 0.115 & 0.188\\
    \end{tabular}
    \caption{Zero-age main sequence luminosities, effective temperatures and central temperatures for primordial monolithic supermassive stars, here defined at the point where nuclear-burning has halted the initial thermal contraction of the star.}
    \label{tab:MS}
\end{table}

\begin{figure}
\begin{center}
\includegraphics[width=0.55\textwidth]{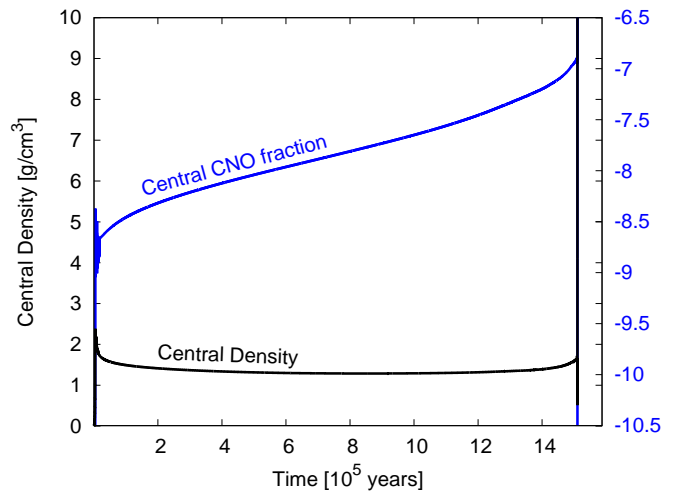}
\caption{Central density (left y-axis) and central CNO mass fraction (right y-axis) as a function of time for a 150,000$\rm{M}_{\odot}$ star over the course of its main sequence lifetime (i.e., until collapse at core hydrogen exhaustion.)}\label{rhoCNO}
\end{center}
\end{figure}

Upon reaching the main sequence, these stars maintain central temperatures of order $10^{8}\,$K, and continue to produce CNO elements via the triple-$\alpha$ reaction.  This increase in catalyzing agents of hydrogen-burning by the CNO-cycle  drives a very gradual expansion of the core as less extreme burning conditions are needed to supply the star's luminosity.  This slowly pushes the central density and the central temperature very slightly to lower values to maintain equilibrium (see Fig. \ref{rhoCNO}), despite the mean molecular weight $\mu$ rising due to the growing helium fraction \citep[see, e.g.,][]{Marigo2001}.  This behaviour of expansion and decrease of core density is unique to Pop III stars and prevents collapse of the star during core hydrogen burning once it has come to thermal equilibrium on the Zero-Age Main Sequence.

We find that all stars born with initial masses below $150\,\kmsun$ survive until core hydrogen is exhausted at the end of the main sequence, whereupon they contract again, risking the post-Newtonian instability once more unless helium-burning reverses the contraction first.  We find this second transition occurs above $\gtrsim 60\,\kmsun$, with larger masses undergoing dynamical collapse at the end of the main sequence and lower masses surviving through to at least onset core helium-burning, and may collapse due to the onset of the post-Newtonian instability or the pair instability at a later evolutionary stage.

Notably, this lower transition occurs immediately above the mass range where, it has previously been claimed, rapid helium-burning may be able to reverse the collapse entirely, allowing stars of $\approx55\,\kmsun$ to undergo thermonuclear explosions \citep{Chen14}. With this in mind, we made a number of trial computations at masses between $150\,\kmsun$--$160\,\kmsun$, in order to search for an additional ``island of explodability'' at the transition from hydrostatic hydrogen-burning to direct collapse.  Such objects are evidently extremely marginal, however, and we were unable to robustly reproduce a similar explosion within this range (i.e., one which did not become a collapse or survive to hydrostatic burning when re-run with a finer resolution in mass grid points and/or time step), despite the extraordinary rapid-proton capture nucleosynthesis immediately prior to collapse. Given the existing uncertainties in, e.g., nuclear reaction rates, one-dimensional modelling of convection and mixing -- especially in such supermassive, extremely radiation-dominated objects -- as well as the strong oscillations in core CNO abundance found after the initial burst of triple-$\alpha$ burning (recall Fig.~\ref{rhoCNO} and discussion above), we do not speculate further at this time, but reserve modeling the final evolution after the onset of collapse, particularly within this mass interval, for a forthcoming study.

\begin{figure}
\begin{center}
\includegraphics[width=0.5\textwidth]{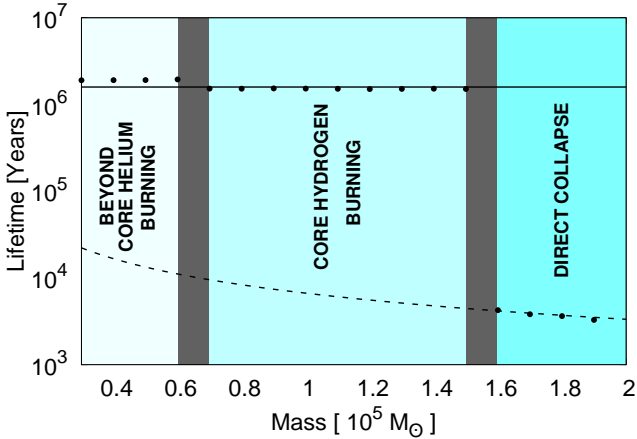}
\caption{Lifetimes of non-rotating, primordial composition monolithic supermassive stars as a function of mass (\textsl{black circles}).  \textsl{Shading colours} indicate the evolutionary state at the time of collapse.  The \textsl{solid black line} denotes the hydrogen-burning lifetime (Eq.~\ref{taunuc}), and the \textsl{dashed black line} denotes the timescale for thermal contraction (Eq.~\ref{tauth}).  The first and second \textsl{vertical dark grey bands} denote the transitions between stars which survive through core-helium burning and those which collapse at the end of core hydrogen-burning; and those which collapse at the end of core-hydrogen-burning and those which never undergo hydrostatic nuclear-burning, respectively.}\label{lifetimes}
\end{center}
\end{figure}

\begin{figure}
\begin{center}
\includegraphics[width = 0.5\textwidth]{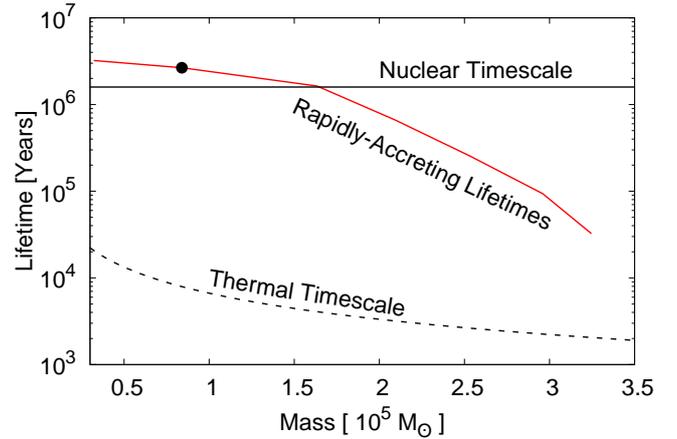}
\caption{The lifetimes of supermassive stars formed through rapid accretion as a function of their final mass at collapse \protect\citep[i.e., for varying accretion rate, see][for more details]{Woods17}.  The large black circle denotes the approximate boundary in this range between objects which collapse while still hydrogen-burning (greater masses) and those which survive to core helium-burning (lower masses).  Note that all previously considered models have undergone a phase of hydrostatic hydrogen-burning prior to collapse -- i.e., for rapidly-accreting objects, no ``truly direct'' collapses have yet been found in numerical simulations.}\label{acclifetimes}
\end{center}
\end{figure}

Figure~\ref{lifetimes} summarises the lifetimes and evolutionary states at collapse of all monolithic supermassive stars as a function of initial mass.  For comparison, Figure~\ref{acclifetimes} shows the lifetimes of supermassive stars which, instead of ``monolithic'' formation, are built up through steady rapid accretion of $0.1$--$10\,\msun\,\yr^{-1}$, e.g., in the atomically-cooled halo scenario, as a function of the \textbf{final} mass which they reach at collapse \citep[from][]{Woods17}.  Two notable differences arise in comparing the fates of supermassive stars in these two extremes:  First, the transition between surviving to core helium-burning and collapsing on the main sequence occurs at higher masses for rapidly-accreting stars (marked with a \textsl{black circle} in Fig.~\ref{acclifetimes} and corresponding to an accretion rate of $\sim0.03\,\msun\,\yr^{-1}$).  This is reasonable, given that a significant portion of the mass in rapidly-accreting models is in a high-entropy envelope, decoupled from the nuclear-burning convective core \citep{Hosokawa13,Woods17,Haemmerle18a}.  Second, there is no transition to truly direct collapses for even the highest accretion rates expected within the atomically-cooled halo scenario, or from a more agnostic standpoint, for infall rates up to $\lesssim10\,\msun\,\yr^{-1}$.  Notably, however, the lifetimes of rapidly-accreting supermassive protostars in our \textsc{kepler} models do appear to trend towards the thermal relaxation timescale at the highest accretion rates.  More rapid accretion rates may be feasible in more exotic massive seed formation scenarios, in particular the mergers of massive, gas-rich proto-galaxies \citep{Mayer10}.  It has been argued \citep{Mayer10,MB18} that under certain conditions such mergers may yield massive ($\sim10^{9}\,\msun$), rotationally-supported disks which may collapse to yield accretion rates of up to $10^{3}$--$10^{5}\,\msun\,\yr^{-1}$.  In a companion study \citep{Haemmerle2019}, we have investigated whether such high accretion rates may allow rapidly-accreting black hole seeds to circumvent hydrostatic hydrogen-burning.  In this case, we found that for accretion rates well above the atomic-cooling regime, central accreting objects cannot maintain hydrostatic equilibrium unless the central protostar has already become supermassive.  We refer the reader to \cite{Haemmerle2019} for more details. 

\section{Discussion}\label{discussion}

Understanding the viability of supermassive stars as a progenitor channel for at least the most massive high-redshift quasars, or even all supermassive black holes, must ultimately await observational confirmation.  In the next decade, next-generation observatories such as the \textit{James Webb Space Telescope} and \textit{Euclid} will, in principle, be able to directly detect supermassive stars themselves at their expected formation redshifts \cite[i.e., $z\sim12$--$20$,  e.g.,][]{Surace18,Surace19}.  The number per unit solid angle per unit redshift which we may expect to be detectable depends linearly on their intrinsic lifetimes ($\sim \Delta t_{\rm{SMS}}$), as well as their formation rate ($\dot{n}_{\rm{SMS}}$):

\begin{equation}
    \frac{\rm{d}N}{\rm{d}z\,\rm{d}\Omega} = \dot{n}_{\rm{SMS}}\,\Delta t_{\rm{SMS}}\,r^{2}\,\frac{\rm{d}r}{\rm{d}z}\label{rate}
\end{equation}

\noindent where $r(z)$ is the comoving distance to redshift $z$,

\begin{equation}
r(z) = \frac{c}{H_0} \int _{0}^{z} \frac{\rm{d}z^\prime}{\sqrt{\Omega_{m}(1 + z^\prime)^3 + \Omega_{\Lambda}}}  
\end{equation}

\noindent Note that in principle both $\dot n_{\rm{SMS}}$ and $\Delta t_{\rm{SMS}}$ will take on distributions of values which may depend on $z$, for a given formation channel and environment. 

\begin{figure*}
\begin{center}
\rotatebox{-90}{\scalebox{0.6}{\input{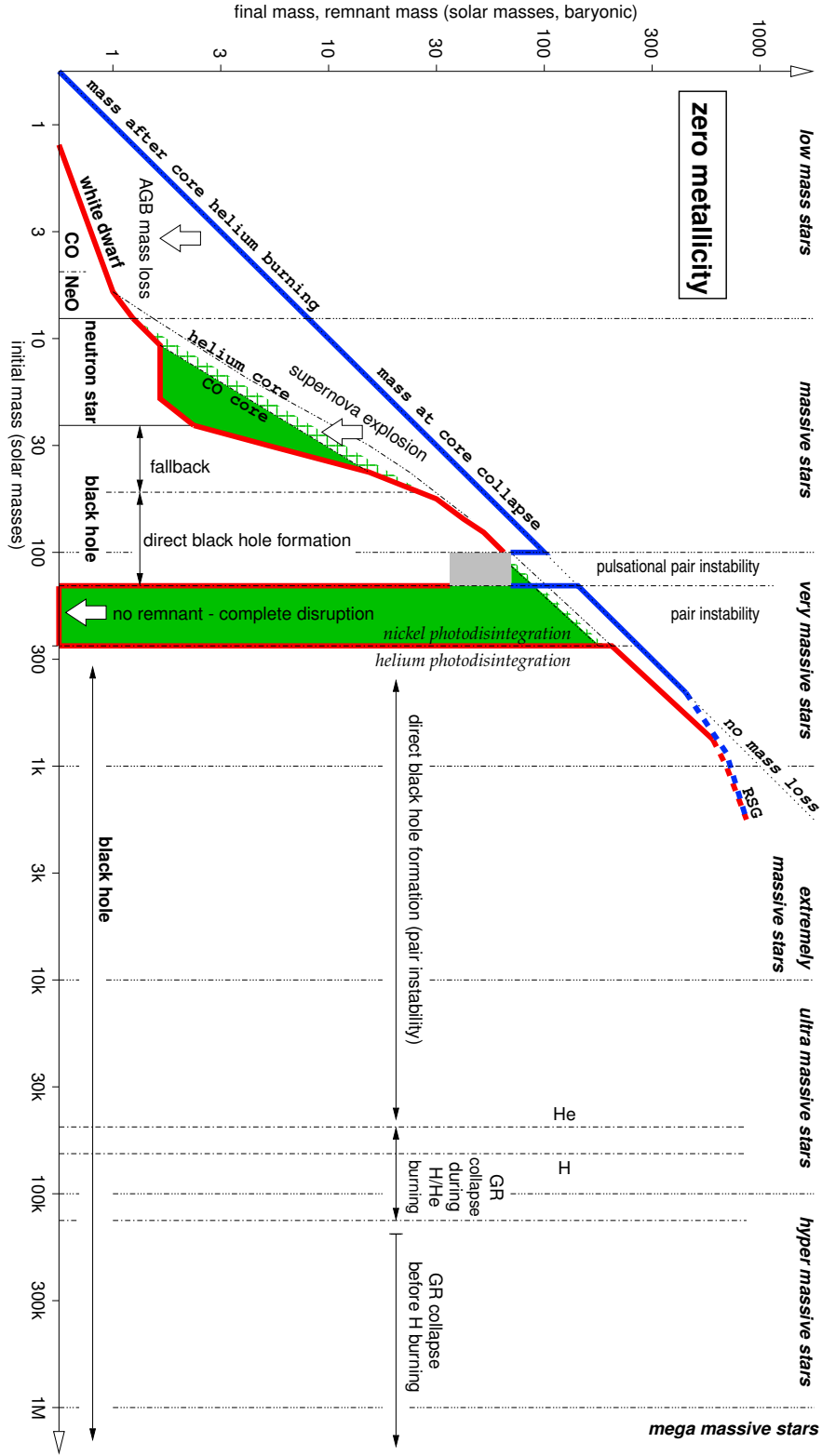}}}
\caption{Extension of the initial-final mass relation for zero-metallicity stars from \protect\cite{Heger02}. For zero metallicity (primordial) stars, final mass at collapse is denoted by the solid blue line (for low-mass stars, we plot the mass at the end of core helium-burning here). The solid red line denotes the final remnant mass after mass loss both during the final stages of stellar evolution as well as explosive mass loss at the time of collapse (with the latter shaded green). ``Conventionally'' massive stars with initial masses greater than $\sim260\,\msun$ collapse directly to black holes due to the e$^{-}$e$^{+}$ pair instability.  This changes in the supermassive regime, as stars with masses greater than $\sim 60\,\kmsun$ collapse at the end of core hydrogen-burning due to the onset of the post-Newtonian Chandraskehar instability, while stars with masses greater than $\sim160\,\kmsun$ reach this instability before the onset of hydrostatic nuclear-burning.}\label{Hegerplot}
\end{center}
\end{figure*}

The continued production of CNO isotopes by the triple alpha process unique to massive Population III stars leads to an expansion during hydrostatic core hydrogen burning and prevents the collapse during core hydrogen burning once the star has reached thermal equilibrium.  The transition at very high masses, and, perhaps, very high accretion rates, from surviving through hydrostatic hydrogen-burning to ``truly direct'' collapse on a thermal timescale carries with it a drop by $2$--$3$ orders of magnitude in expected lifetime prior to black hole formation.  Therefore, understanding not only the formation rates but the total masses of supermassive objects in any given scenario is critical to evaluating future prospects for detection.  For the collapse of dense stellar clusters, the loss of a substantial fraction of such a cluster's mass by ejection of stars via 3-body interactions is thought to strongly limit the total mass of any central supermassive star formed by this channel, with an upper bound potentially as low as $\sim 3\,\kmsun$ \citep[e.g.,][]{Devecchi12}.  This is still well within the mass regime where collapse is expected due to the pair-instability at the end of the star's nuclear-burning lifetime (Fig.~\ref{Hegerplot}).  In this case, no ``truly direct'' (or ``dark'') collapse is possible within dense stellar clusters, even within the idealized model presented here.

Recently, it has been suggested that the photospheres of growing supermassive stars embedded within dense stellar clusters may greatly inflate, in analogy with supermassive stars which grow by rapid accretion in the atomically-cooled halo scenario.  The resulting larger cross-section for mergers may then be able to boost the possible final mass of the central object formed by cluster collapse by an order of magnitude or more \citep{Boekholt18}.  In this case, however, the structure of the star would similarly be more analogous to that of rapidly-accreting supermassive stars, and we would still expect a phase of hydrostatic nuclear-burning prior to collapse \citep{Woods17}.  Given our present understanding of supermassive stellar evolution at accretion rates accessible by extreme environments in the early Universe, then, it appears an initial phase of hydrostatic nuclear-burning is inescapable prior to collapse, putting a firm lower bound on the lifetimes of supermassive ``direct collapse'' black hole seeds of $\sim30\,\kyr$--$100\,\kyr$ for the most extreme accretion rates possible in the atomic-cooling halo regime, with lifetimes of $\sim1$--$2$ million years being more typical for lower accretion rates or in the collapse of dense stellar clusters. Only for very extreme accretion rates ($\gg 10\,\msun\,\yr^{-1}$), arising from, e.g., the merger of massive protogalaxies \citep{Mayer10}, may ``truly direct'' collapses be possible, however, this requires further study \citep{Haemmerle2019}.

Aside from the direct detection of supermassive stars, what other prospects exists for observational constraints?  If there is significant mass return from supermassive stars, either during or at the end of their lives, this may leave a distinct nucleosynthetic signature that can be sought in the local Universe.  Notably, it has been suggested that eruptive mass loss from supermassive stars during their hydrogen-burning main sequence may be able to account for the abundance anomalies in proton-rich elements found in globular clusters \citep{Denissenkov14,Gieles18}.  This is due to the very high central temperatures at which hydrogen-burning takes place in these stars, although a mechanism for producing such eruptions has proven elusive. On a speculative note, one possibility may be the rapid nuclear-burning of a sudden influx of hydrogen ingested during a merger with a more typical Pop III star; for our $150\,\kmsun$ star, the total binding energy near the end of the main sequence is approximately $3\times 10^{54}\,\mathrm{erg}$, equivalent to the energy released from nuclear-burning of all the hydrogen in a $\sim300\,\msun$ star.  Whether merger events are able to lead to eruptive episodes that partially or fully unbind the star requires considerable further investigation.  In the event of an absence of additional ``islands of explodability,'' such as the narrow range of masses found for helium-burning supermassive stars by \cite{Chen14}, evidence for the mass distribution of supermassive stars, and their final fates, may need to be inferred from the results of future efforts to find intermediate mass black holes \citep{Mezcua17,Koliopanos17,Woods19}, massive black hole seeds accreting at high redshift \citep[e.g.,][]{Pacucci19}, and critically, direct detections of supermassive stars themselves \citep{Surace18,Surace19}.

In the meantime, our results for the evolution of monolithic stars are summarized in the context of more typical stellar evolution in Fig.~\ref{Hegerplot}. Based on these results, we may place a tentative upper bound on the masses of stars which collapse at the end of their nuclear-burning lives upon encountering the pair instability, with primordial stars more massive than $\sim60\,\kmsun$ collapsing at the end of the hydrogen-burning main sequence due to the post-Newtonian Chandrasekhar instability.  This provides a physically-motivated division of any very massive primordial star from the truly ``supermassive'' regime, at least for the non-rotating case. Furthermore, we may place a similar absolute limit of $\sim150\,\kmsun$ on the total stellar mass of a primordial, thermally-relaxed, non-rotating star, for which a hydrogen-burning main sequence is possible.  We conclude that for the non-rotating, monolithically-formed case, no bound object of primordial composition with mass greater than $\sim150\,\kmsun$ may be truly considered a ``star.''

\section*{Acknowledgements}

We would like to thank Ralf Klessen and Daniel Whalen for valuable discussions, as well as Emily Troup, Brian Crosby, and Ryan Poitra for their early investigations of this problem as summer students.  TEW acknowledges support from the NRC-Canada Plaskett fellowship.  AH has been supported by a grant from the Science and Technology Commission of Shanghai Municipality (Grants No.\ 16DZ2260200) and National Natural Science Foundation of China (Grants No.\ 11655002) and  benefited from support by the National Science Foundation under Grant No. PHY-1430152 (JINA Center for the Evolution of the Elements).
LH was sponsored by the Swiss National Science Foundation
(project number 200020-172505).




\bibliographystyle{mnras}
\bibliography{refs} 


\bsp	
\label{lastpage}
\end{document}